\documentclass[10pt,letterpaper]{article}
\usepackage{opex3,amsmath}
\usepackage{cite}
\usepackage{subfigure}

\begin{document}

\title{Metropolitan all-pass and inter-city quantum communication network}
\author{Teng-Yun Chen$^{1,2}$, Jian Wang$^{1,2}$, Hao Liang$^{1}$, Wei-Yue Liu$^{2,3}$, Yang Liu$^{1,2}$,
Xiao Jiang$^{1,2}$, Yuan Wang$^{1}$, Xu Wan$^{1}$, Wei-Qi Cai$^{1}$, Lei Ju$^{1,2}$, Luo-Kan Chen$^{1,2}$, Liu-Jun Wang$^{1}$, Yuan Gao$^{1}$,
Kai Chen$^{1}$, Cheng-Zhi Peng$^{1}$, Zeng-Bing Chen$^{1}$,and Jian-Wei Pan$^{1}$}


\begin{abstract}
We have demonstrated a metropolitan all-pass quantum communication
network in field fiber for four nodes. Any two nodes of them can be
connected in the network to perform quantum key distribution (QKD).
An optical switching module is presented that enables arbitrary
2-connectivity among output ports. Integrated QKD terminals are
worked out, which can operate either as a transmitter, a receiver,
or even both at the same time. Furthermore, an additional link in
another city of 60 km fiber (up to 130 km) is seamless integrated
into this network based on a trusted relay architecture. On all the
links, we have implemented protocol of decoy state scheme. All of
necessary electrical hardware, synchronization, feedback control,
network software, execution of QKD protocols are made by tailored
designing, which allow a completely automatical and stable running.
Our system has been put into operation in Hefei in August 2009, and publicly
demonstrated during an evaluation conference on quantum network
organized by the Chinese Academy of Sciences on August 29, 2009.
Real-time voice telephone with one-time pad encoding between any two of
the five nodes (four all-pass nodes plus one additional node through relay)
is successfully established in the network within 60km.
\end{abstract}

\address{$^1$Hefei National Laboratory for Physical Sciences at Microscale
and Department of Modern Physics, University of Science and
Technology of China, Hefei, Anhui 230026, China\\
$^2$Anhui Quantum Communication Technology Co., Ltd., Hefei,
Anhui 230088, China\\
$^3$School of Information Science and Engineering, Ningbo
University, Ningbo, Zhejiang 315211, China}

\ocis{(270.0270) Quantum optics; (060.0060) Fiber optics and optical
communications; (060.5565) Quantum communications.}


\section{Introduction}

It has been nearly 3 decades since proposal of quantum key distribution (QKD) \cite{BB84}. There are significant
theoretical developments and experimental schemes demonstrated, to name a few of them, see, e.g.,
\cite{Gisin1997,Nishioka2002,Grosshans2005,GYS2004,Peng2005,Chen2006,Honjo2006,superconducting,Qiang08}, which
enables rapid developments for point to point (PTP) key establishment. Compared with classical optical
communications, it is very critical to realize secure connections beyond PTP links by employing QKD, which would
offer very promising network applications.

Currently there are mainly two topology structures to expand existing QKD links. One is to use so-called trusted
relay architecture. Another one is to implement a transparent network architecture, similar to the case of
classical optical network. By using trusted relay, one can increase communication distance for QKD arbitrarily. Also
different type of QKD links are compatible in such architecture. However, one should ensure privacy for the relay
sites to guarantee security for the whole network. The European SECOQC (Secure Communication based on Quantum
Cryptography) quantum network \cite{SECOQC} has utilized this kind of paradigm. We have also demonstrated a 3-node
communication network based on a trusted relay with each adjacent link of about 20km \cite{oetyc2009}. By employing
optical switching techniques, on the other hand, one can achieve low network complexity for constructing
transparent connections. Unfortunately, the application of switching solely can not increase communication distance
and key generation rate for QKD. A practical way is to combine additional trust relays to construct a hybrid
scalable network. There are many ways to maintain transparent network implementations, such as via optical switching
\cite{Elliott2002,Elliott2005,NIST-net}, passive optical splitting \cite{Phoenix1995,Townsend1997}, or
wavelength-division multiplexing (WDM) \cite{WDM1997,Guo2009}. By using beam splitters, one cannot choose
connections freely in a passive network, whereas in a WDM-type network, the transmitter has to prepare in advance
laser diodes for operating in corresponding wavelength and choosing communications sites. Optical switching has
been extensively used such as in the DARPA network\cite{Elliott2002,Elliott2005}, the 3-node NIST network
\cite{NIST-net}, in \cite{JST-net} together with a relay, and in \cite{LosAlamos-net} for a dynamically
reconfigured network.

We have complemented and improved performance of QKD network over
existing demonstrations in several aspects. This mainly includes
all-pass optical switching, novel QKD terminals, 60$\sim$130 km
inter-city links, with the help of scalable hybrid network topology.
The earlier demonstrations for optical switching network used single
optical switch \cite{Elliott2005,NIST-net} and achieved only one-way
connection for some of communication parties. We have managed to
design an optical switching equipment that enables 2-connectivity
for any input and output ports. Therefore arbitrary interconnection
can be created between any two of input and output ports. An
all-pass metropolitan QKD network with star-type topology is then
accomplished in Hefei city of China. The QKD communications are
carried out for any two nodes based on polarization coding.
Real-time voice encryption and decryptions are successfully
demonstrated through one-time pad coding. Besides, each of the nodes
could work as a transmitter or a receiver, or even as both
altogether in field experiment. Our design thus offers a potential
for duplex communication for the node terminals. Moreover, with such
switching equipments and terminals, it is straightforward to
construct a network with routing function for quantum signals,
similar to the case of classical communications. In our case, the
distribution of quantum channel is automatically decided base on
communication request, loss of the backbone network, busy or on-off
status for fibers. An additional node in Feixi county that is about
60 km fiber distance far from relay node at USTC site in Hefei city,
is further added to achieve an inter-city QKD network. With help of
trusted relay architecture, we have again established real-time
voice communication between this node and any node from the
metropolitan network. The network's robustness is verified for this
5-node hybrid structure, by moving further the additional node to
Tongcheng city that is about 130 km for fiber distance far from USTC
site (the actual distance is about 100km). It works well to
communicate with any node from metropolitan network. The network
deployment took place in August 2009, and was publicly demonstrated
during an evaluation conference on quantum network organized by the
Chinese Academy of Sciences on August 29, 2009. Our hybrid network
has realized all necessary functions for a practical QKD network by
developing and integrating several key modules including active
optical switching for connecting any two ports, hybrid network
architecture by integrating optical switching and trusted relay,
tailored QKD network processing and control software, automated
distribution for quantum channels, intercorporate communication
functions of transmitter and receiver for QKD terminals in every
node of all-pass network, guaranteed security by employing decoy
state schemes
\cite{Hwang2003,Wang2005,Lo-Ma-Chen2005,Wang2005PRA,Lo2005PRA}. Thus
we hope that our demonstration could provide a critical step towards
practical QKD network in a large scale.

\section{Optical switching and network architecture}

In order to achieve an all-pass network in a metropolitan network, one of the key ingredient is an equipment that
allows arbitrary interconnecting for fibers connecting with different communication parties. The earlier uses for
optical switching network \cite{Elliott2005,NIST-net} has not yet operated with this function. In the
demonstration, we have managed to produce an 8-port optical switching equipment that allows interconnecting of any
two ports. A star-type network is accomplished by connecting every node to our optical switching equipment.
Compared with other types of network, there is no special requirements for all-pass network with optical switches.
Particularly, this network offers distribution of quantum channel with relatively low loss and enjoying the
advantage of easy controlling with classical network commands. We use the mechanical optical switches, which
provides high degree of isolations among different ports without direct connections. Moreover, there is no induced
additional noise when all the ports work at the same time. Furthermore, this kind of optical switching techniques
provides standard single mode fiber channel, which enables hybrid connections for different schemes of QKD and
holds potential performance improvement if combined with WDM techniques developed in \cite{WDM1997,Guo2009}.

\begin{figure}[ptb]
\begin{center}
\includegraphics[width=95mm]{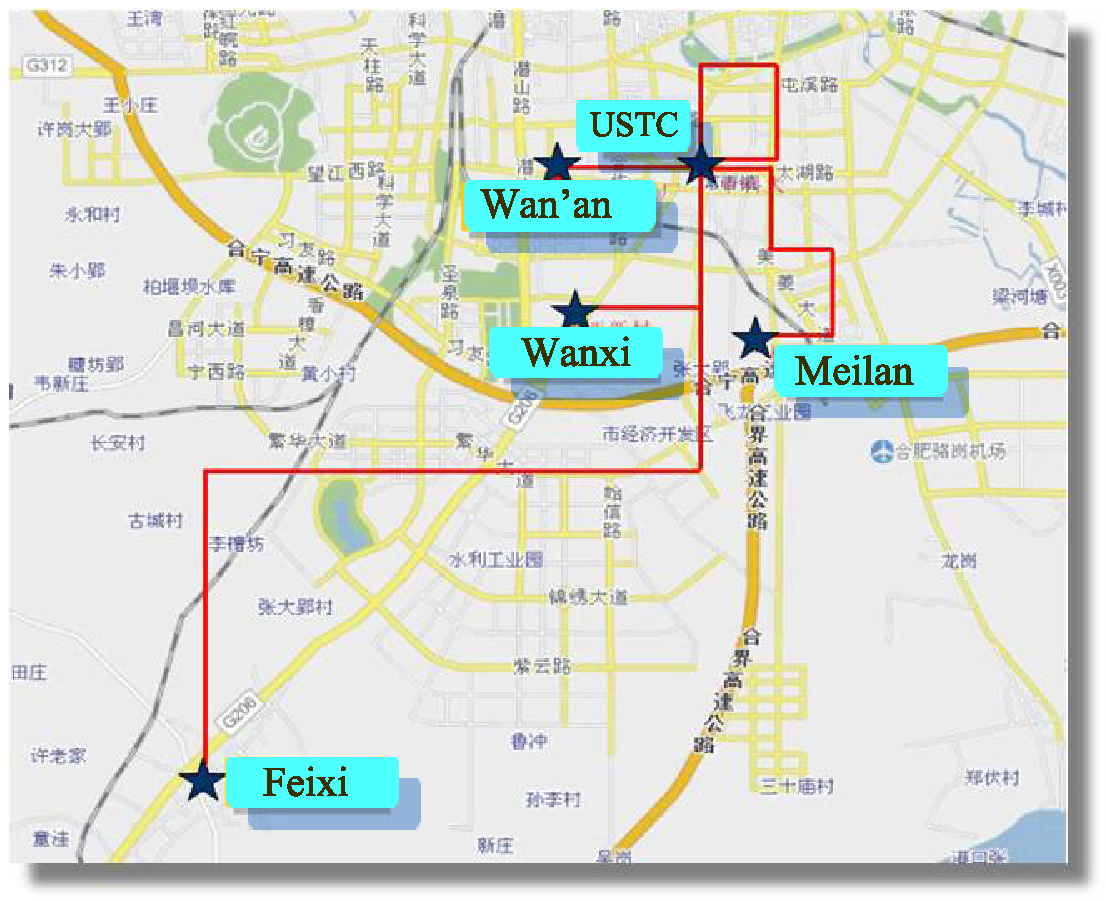}
\end{center}
\caption{Metropolitan all-pass quantum communication network constitutes 4 nodes including USTC, Wan'an, Meilan, and
Wanxi. A circle link of 10 km goes back USTC through underground optic fiber cable, which is used to simulate a
separately remote node. An additional node in Feixi county (finally moved to Tongcheng city) that is about 60 km fiber
distance (130 km fiber distance) far from USTC site in Hefei city, is further connected to achieve an inter-city QKD network, when the USTC node
serves as a trusted relay.}
\label{inter-city}
\end{figure}

It is neither proved nor quantified for unconditional security
through trusted relay architecture up to now. We remark that trusted relay
architecture is however a very practical architecture, which could
in principle extend range of secure distance for QKD in a large
scale. Once the relay nodes are secure, any other nodes can
communicate securely with the help of relay nodes. In addition, the
relay node provide an interface allowing interoperability of
heterogeneous QKD devices, which expand practical applications of
QKD network. We have set up a star-type network based on a trusted
relay in USTC site, as shown in Fig.~\ref{inter-city}. If combined
with multi-level optical switching and multiple trusted relays, this
network topology is scalable to arbitrarily expand to be a complex
network with additional QKD devices. We have set up two types of QKD
terminals. The terminals connected to the optical switching module
are those equipped with functions of being both transmitter and
receiver. The USTC node connected to optical switching has acted as
a trusted relay, one part of which has been equipped with such a
terminal. Another part of USTC node is treated as a receiver, and
uses high speed system that we have developed in \cite{200kmQKD}
without amplification of synchronized signal in between link. All
the nodes are running with the standard BB84 protocols base on decoy
state schemes \cite{Hwang2003,Wang2005,Lo-Ma-Chen2005,Wang2005PRA,Lo2005PRA}.
The performances are tested and verified for all of possible connections
among the 5 nodes.

\section{QKD terminal devices for network applications}

As shown in Fig.~\ref{inter-city}, the star-type network provides quantum channels for any two nodes among the
4 nodes in the metropolitan area of Hefei city. To run the network, one has to update the normal PTP QKD links
to cover all the nodes. Considering the fact that the arbitrary connection should be possible for any two of the nodes, we have made a integrated design for terminals that
could work either a transmitter or a receiver.

We make use of weak coherent states coming from laser diodes as optical source. Decoy state scheme
is implemented for all the links, to extend significantly secure distance and improve key generation rate with
proved security. The main idea for decoy scheme is to insert randomly decoy states with different intensity
from the signal state during the transmission process. In the receiver side, through detection rates and
quantum bit error rates for both the signal states and the decoy states, one can analyze to derive maximum possible
information leaked to eavesdropper. Thus two communications sides could then generate secure keys after error corrections
and privacy amplification process.

\begin{figure}[ptb]
\begin{center}
\includegraphics[width=100mm]{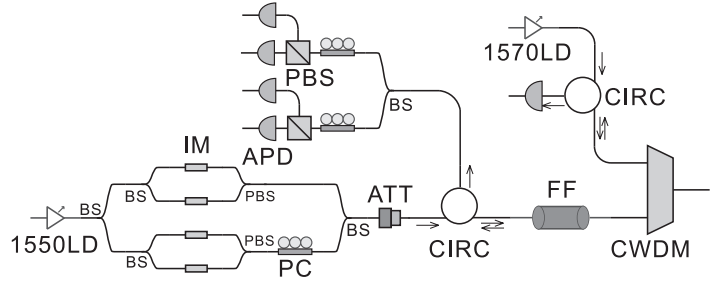}
\end{center}
\caption{Schematic view of a QKD device terminal in the experimental setup. When the device serves as a
transmitter, four polarizing states are generated by a 1550 nm laser diode after two cascaded BS with additional
intensity modulation, before combined by another output BS. Signal and decoy states are also controlled at this
stage by random choice. After suitable attenuation, the optical pulses go through a circulator before combining in
a CWDM with synchronization pulse. When the terminal serves as a receiver, the optical pulses enter at CWDM, and then
decode quantum signals out, for going through circulator at the detection part. Synchronization signal goes along
another circulator for clock signal detection. Here, LD: Laser Diode; IM: Intensity Modulator; PBS: Polarization
Beam Splitter; PC: Polarization Controller; ATT: Attenuator; CIRC: Circulator; FF: Fiber Filter; CWDM: Coarse
Wavelength-division multiplexing.}
\label{qkdterminal}
\end{figure}

A schematic view of QKD terminal in every terminal of metropolitan network is illustrated in Fig.~\ref{qkdterminal}.
In order to maintain a relatively high key generation rate, we set the photon number intensity to be
$0.6:0.2:0$/pulse for signal states, decoy states and vacuum states, respectively. The occupancy proportion is set
as $6:1:1$. When the terminal serves a transmitter, the optical pulses are modulated randomly with
intensity ratios of 3:1:0 with two cascaded intensity modulators after 1550 nm laser diodes to generate
the three kinds of states. Four types of polarized states of $H/V/+/-$ are prepared also at this stage
after two cascaded beamsplitters (BS) to represent horizontal, vertical $+45^\circ$ and $-45^\circ$ polarization states.
With additional attenuations, the pulses are outputted to field fiber after adjusting the average photon number
intensity to be $0.6/$pulse for signal states. In the receiver side, standard BB84 detection scheme is applied.
When the terminal serve as a receiver, in the detection part the input optical pulses are divided by a beamsplitter
(BS) into two arms corresponding to $H/V$ and $+/-$ basis detection unit, respectively. In each arm, there is a
polarization controller that will actively compensate possible polarization displacement in fiber channel. The
polarization beamsplitters (PBS) are then for choosing $H/V$ or $+/-$ measurement basis before the pulses entering
detectors.

To incorporating the transmitter function in the same terminal, we have used fiber circulator to isolate the input
pulses from field fiber, and the output pulses from the optical source.  Whenever the terminal plays a role of transmitter,
pulses from laser diode will go through the circulator to outer fiber. Whenever the terminal works as a receiver, the
input pulses from outer fiber goes over the circulator and then arrives at detectors. To avoid strong reflection
caused by Rayleigh backward scattering for light other than quantum signals in WDM output ports, we have added a
narrowband ($\sim$ 100GHz) fiber filter after the circulator with central wavelength of 1510.12 nm.  This improvement
could dramatically reduce unwanted disturbance caused by noisy light (mainly from synchronization light), and thus
contribute low quantum bit error rate (QBER). Moreover the setting will result in a high visibility for outputs of
polarizing quantum states. As the detectors are working in gate mode, they require external trigger
signals. We have therefore use strong optical pulses as synchronization signal, which allow detections by normal
photoelectrical diodes. Whenever there is a quantum pulse is emitted, there is a synchronization pulse from
synchronization laser of 200 nw, to tag timing information. For purpose of reducing cost for using fiber, we have
managed to use 1570 nm optical pulses as synchronization signal. Together with quantum optical pulses, they are
combined into a coarse wavelength division multiplexing module for outputting in field fiber. Another advantage is that
the light with wavelength 1570 nm contributes relatively small anti-Stokes scattering for light of 1550 nm,
which reflects very low disturbance for quantum signals.

\section{Performance in field fiber}
Our QKD network is base on the running field fibers of China Netcom Group Corp Ltd. All of the four nodes are
connected to the optical switching module at laboratory situated in USTC. It should be remarked that there is one
circle link of 10 km goes back USTC through underground optic fiber cable, to simulate a separately remote node.
The fifth node in Feixi county that is 60 km fiber distance far from USTC is connected to all-pass metropolitan network in Hefei city,
with USTC site acting as a trusted relay. The specification parameter for
all the links are shown in Table \ref{table1}.

\begin{table}[tbh]
\caption{Measured specifications for QKD network}
\centering
\begin{tabular}[b]{llllll}
\hline\hline
link       & Circle link USTC & Wan'an   & Meilan   & Wanxi    & Feixi \\ \hline
Distance   & 10.047 km        & 8.447 km & 9.904 km & 8.417km  & 60km  \\
Fiber loss & 2.82 dB          & 2.65 dB  & 2.86 dB  & 2.75 dB  & 17 dB \\
\hline\hline
\end{tabular}
\label{table1}
\end{table}

All of the four nodes including USTC, Wan'an, Wanxi and Meilan utilize integrated QKD terminals with functions of
transmitter and receiver. The repetition rate of 4 MHz is used for the laser source at these nodes. As mentioned
before, the average photon number is 0.6 and 0.2, for sinal states and decoy states, respectively. The receivers
use single photon detectors of InGaAs type with two id201 detectors from id Quantique and two detectors produce by
East China Normal University. The detectors' efficiency is about 10\% for all of four detectors in each node. Once
powered on, our system can automatically execute whole adjusting and feedback process, wait connecting request from
any nodes, and choose their corresponding working modes. We find that the systems can work perfectly with current
commercially available underground fiber cables. After extensive monitoring and tests, the system has proven very
robust and stable, with a consistent key generation rate for a period of 24 h.

For the whole network links, we have measured and derived all the relevant parameters, as listed in Table
\ref{table2}.
\begin{table}[ht]
\caption{Measured and derived specifications for quantum network based on decoy states}
\centering
\begin{tabular}[b]{llllllll}
\hline\hline
Para.     & Meilan-USTC & USTC-Meilan  & USTC-Wanxi   & Wanxi-USTC   & Wanxi-Wan'an \\ \hline
Sifted R  & 11.0k       & 9.74k        & 10.0k        & 8.02k        & 8.00k        \\
Final  R  & 1.45k       & 1.20k        & 1.95k        & 1.45k        & 1.30k        \\
$E_{\mu}$ & 1.58\%      & 1.47\%       & 1.51\%       & 1.53\%       & 1.35\%       \\
$E_{\nu}$ & 4.00\%      & 4.10\%       & 4.99\%       & 4.99\%       & 4.41\%       \\
$Q_{\mu}$ & $8.21\times 10^{-3}$    & $5.81\times 10^{-3}$     & $7.25\times 10^{-3}$     & $5.83\times 10^{-3}$     & $7.15\times 10^{-3}$     \\
$Q_{\nu}$ & $2.71\times 10^{-3}$    & $1.90\times 10^{-3}$     & $2.32\times 10^{-3}$     & $1.91\times 10^{-3}$     & $2.20\times 10^{-3}$     \\
$Y_{0}$   & $2.03\times 10^{-4}$    & $1.38\times 10^{-4}$     & $2.04\times 10^{-4}$     & $1.70\times 10^{-4}$     & $1.78\times 10^{-4}$     \\ 
\\
\hline\hline
Para.     & Wan'an-USTC &Meilan-Wanxi &Meilan-Wan'an & USTC-Wan'an  & Wanxi-Meilan  \\ \hline
Sifted R  & 8.33k       & 8.54k       & 9.39k        & 8.17k        & 7.97k         \\
Final  R  & 1.40k       & 1.43k       & 2.54k        & 1.82k        & 1.75k         \\
$E_{\mu}$ & 1.67\%      & 1.70\%      & 1.28\%       & 1.43\%       & 1.68\%        \\
$E_{\nu}$ & 5.40\%      & 4.43\%      & 3.48\%       & 2.79\%       & 4.28\%        \\
$Q_{\mu}$ & $5.80\times 10^{-3}$    & $6.79\times 10^{-3}$    & $6.86\times 10^{-3}$     & $7.33\times 10^{-3}$     & $6.23\times 10^{-3}$      \\
$Q_{\nu}$ & $1.90\times 10^{-3}$    & $2.30\times 10^{-3}$    & $2.29\times 10^{-3}$     & $2.48\times 10^{-3}$     & $2.21\times 10^{-3}$      \\
$Y_{0}$   & $1.75\times 10^{-4}$    & $1.86\times 10^{-4}$    & $1.19\times 10^{-4}$     & $1.33\times 10^{-4}$     & $1.58\times 10^{-4}$      \\ 
\\
\hline\hline
Para.     &Wan'an-Meilan & Wan'an-Wanxi& Feixi-USTC    \\ \hline
Sifted R  & 7.33k        & 8.39k       & 18.0k         \\
Final  R  & 1.40k        & 1.21k       & 4.50k         \\
$E_{\mu}$ & 1.60\%       & 1.56\%      & 1.13\%        \\
$E_{\nu}$ & 5.16\%       & 4.97\%      & 1.71\%        \\
$Q_{\mu}$ & $6.43\times 10^{-3}$     & $5.68\times 10^{-3}$    & $1.64\times 10^{-4}$      \\
$Q_{\nu}$ & $2.16\times 10^{-3}$     & $1.91\times 10^{-3}$    & $6.60\times 10^{-5}$      \\
$Y_{0}$   & $1.77\times 10^{-4}$     & $1.74\times 10^{-4}$    & $1.13\times 10^{-6}$      \\ \hline\hline
\end{tabular}
\label{table2}
\end{table}
Here we use post data processing method followed from results of
\cite{GLLP} and \cite{Hwang2003,Wang2005,Lo-Ma-Chen2005,Wang2005PRA,Lo2005PRA}.
The key generation rate that can be achieved is as follows
\begin{equation}
R\geq q\{-Q_\mu f(E_\mu) H_2(E_\mu)+Q_1[1-H_2(e_1)]\},
\label{R}
\end{equation}
where the subscript $\mu$ is the average photon number per signal in
signal states. For convenience, we denote $\nu$ the average photon number
per pulse for decoy state.  $Q_\mu$ and $E_\mu$ are the measured gain and the
QBER for signal states, respectively; $q$ is an
efficiency factor for the protocol. $Q_1$
and $e_1$ are the unknown gain and the error rate of the true single
photon state in signal states. The decoy state method can estimate the lower bound
of $Q_1$ denoting as $Q_{1}^{L}$, and the upper bound of $e_1$
denoting as $e_{1}^U$, and then one can achieve maximum possible secure key
rate. The $H_2(x)$ is the binary entropy function:
$H_2(x)=-x\log_2(x)-(1-x)\log_2(1-x)$, while the factor $f(x)$ is for considering
an efficiency of the bi-directional error correction
\cite{Brassard-Salvail1994}.

We follow here the method developed in \cite{GLLP} and
\cite{Hwang2003,Wang2005,Lo-Ma-Chen2005,Wang2005PRA,Lo2005PRA}
to estimate good bounds for
$Q_1$ and $e_1$. After experimentally measuring all the relevant parameters as listed in Table \ref{table2}, we can
input the following bounds for calculating final key generation rate \cite{Lo2005PRA}
\begin{eqnarray}
Q_1 \ge Q_1^L &=& \frac{\mu^2e^{-\mu}}{\mu\nu-\nu^2}(Q_\nu^L
e^{\nu}-Q_\mu e^\mu\frac{\nu^2}{\mu^2}-Y_0^U
\frac{\mu^2-\nu^2}{\mu^2}),\\
e_1 \le e_1^U &=& \frac{E_\mu Q_\mu-Y_0^Le^{-\mu}/2}{Q_1^{L}},
\label{bounds}
\end{eqnarray}
in which
\begin{eqnarray*}
Q_\nu^L &=& Q_\nu(1-\frac{10}{\sqrt{N_\nu Q_\nu}}),\\
Y_0^L &=& Y_0(1-\frac{10}{\sqrt{N_0 Y_0}}),\\
Y_0^U &=& Y_0(1+\frac{10}{\sqrt{N_0 Y_0}}),
\label{pbounds}
\end{eqnarray*}
Here $N_\nu$, and $N_0$ are numbers of pulses used as decoy state and vacuum state, respectively, while $Q_\nu$ is
the measured gain for the decoy states. The measured counting rate for vacuum decoy state is denoted by $Y_0$.

From Table \ref{table2}, we see a final key rate of more than 1.2 kbps whenever QBER is less than 2\%, for a
typical running of 400 s for our system. This has already excluded 1/5 period consuming for adaptive feedback control.
We have estimated the bounds for $Q_{1}^{L}$ and $e_{1}^U$ by considering the statistical fluctuations for vacuum
states, gains for signal states and decoy states within 10 standard deviations, which
ensure the final keys rates promises a confidence interval of about $1-1.5\times 10^{-23}$.
Although the distance is relatively long for Feixi-USTC link, we obtain highly fast key rates of 4.5 kbps. This is
mainly due to several essential elements that we have maintained. We have managed to achieve 320 MHz high repetition
rate for optical source. With the help of superconduction detectors, extremely low dark counts or counts from
unwanted light is anticipated. It is therefore attained for high detection counting rate of about 10 kbps for each
arm, and low QBER of typically less than 1\%. Based on these resource of secure keys, we have finally tested in
application layer to realize voice communication with one-time pad. The voice communication is not only implemented
in all of the four nodes in the metropolitan network, each of which has also successfully
created secretly audio communication with Feixi node. We have attempted to move the system in Feixi node to Tongcheng city
that is about 130 km from USTC, with an approximate fiber loss of 29 dB for transmission.
Again we have demonstrated successful operation of QKD, by using
broadband wireless network from China Telecom as classical communication channel.
A final key rate around 0.2 kbps and QBER of less than 2 \% are achieved.

\section{Conclusions and perspectives}
We have demonstrated an all-pass quantum communication network in
field environment. Hybrid network architecture is illustrated to
construct an inter-city network by combining the metropolitan
quantum network and a trusted relay, which is capable of extending
reach of network nodes arbitrarily. All of necessary functions and
equipments are realized, including seamless integrating of all-pass
optical switching, trusted relay, decoy state protocol, tailored QKD
network hardware, and software control etc. Integrated QKD terminals
are developed, which can operate both as a transmitter or a receiver
with automated switching. The designed terminals are in fact
possible to be used as both transmitter and receiver at the same
time, which will double key generation rate with suitable software
and electrical control hardware. The hybrid architecture by using of
all-pass structure and trusted relay would enables a scalable
network for arbitrary distance. The results reported in this paper
would help to make a significant step for a practical QKD network in
widespread implementation and associated applications.

\section*{Acknowledgments}
We acknowledge the financial support from the CAS, the National
Fundamental Research Program of China under Grant No.2006CB921900,
the National High Technology Research and Development Program (863
Program) of China, the NNSFC and the Fundamental Research Funds for
the Central Universities.
\end{document}